\documentstyle[12pt]{article}
\begin{document}

\thispagestyle{empty}

\hspace*{9cm}  PDMI PREPRINT - 1/1995
\vspace*{35mm}
\begin{center}
{\LARGE Discrete Heisenberg-Weyl Group and Modular Group}
\par\vspace*{22mm}\par
{\large L.~D.~Faddeev}
\par\bigskip\par\medskip
{\em St.Petersburg Branch of Steklov Mathematical Institute,\par
Fontanka 27, St.Petersburg 191011,
Russia} \par\medskip
{\em Research Institute for Theoretical Physics,
University of Helsinki,\par Siltavuorenpenger 20C,
SF 00014 Helsinki, Finland}
\end{center}
\vspace*{15mm}

\begin{abstract}
It is shown that the generators of two discrete Heisenberg-Weyl groups
with irrational rotation numbers $\theta$ and  $-1/ \theta$ generate
the whole algebra $\cal B$ of bounded operators on $L_2(\bf R)$.
The natural action of the modular group in $\cal B$ is implied.
Applications to dynamical algebras appearing in lattice regularization
and some duality principles are discussed.
\end{abstract}

\newpage

Writing a contribution to a memorial volume one always feels the mixture
of the admiration for passed away and sorrow of a loss. Because of
a difference in age and geographical (and/or political) obstructions
I did not have any significant personal scientific encounters with
J.Schwinger.  But our brief exchanges during several short meetings
and most of all reading his papers influenced my way of thinking and
writing to a great extent.

Together with his colossal work on QED and general quantum field theory,
J.Schwinger did pay attention to technical problem, pertaining to the
ordinary nonrelativistic quantum mechanics \cite{1}.  So I hope, that
the comments in this paper, which are confined to the objects  in a
simplest Hilbert space of quantum theory, namely $L_2 (\bf R)$,
still would amuse him.  My excuse is that these comments were generated
by a dynamical problem in quantum field theory.

What I want to discuss is to my belief the base of several "duality
principles", which appear nowadays, i.e., in connection with string
theory and conformal field theory. However I shall not go into
technicalities of these subjects. More on this wll be mentioned in
the conclusion of this paper.

I consider a Hilbert spase ${\cal H}$, where  the usual coordinate and
momentum operators $Q$ and $P$, satisfying the Heisenberg commutation
rule

\[   [P, Q] = -i \hbar I          \]
are irreducibly represented. For example, we can take
${\cal H} = L_2 (\bf R)$  with elements $\psi (x)$ and usual action
of $Q$ and $P$
\[ P \psi (x) = \frac{\hbar}{i} \frac{d}{dx} \psi (x); \;\;\;\;\;
Q \psi (x) = x \psi (x) \]
(coordinate representation).

Together with $Q$ and $P$ consider their exponentials
\[ u = e^{iPq/\hbar},  \;\;\;  v = e^{iQp/\hbar}    \]
with the Weyl commutation relations
\[ uv = e^{2\pi i \theta} vu,  \]
where
\[ \theta =  \frac{pq}{2\pi \hbar}.  \]

Here $p$ and $q$ are fixed real numbers with dimensions of momentum and
coordinate, respectively, so that $\theta$ is dimensionless.
I could of course fix the units in such a way, that the  dimensions
disappear. Only the parameter $\theta$ is essential in what follows.

It is clear that (for fixed $p$ and $q$) the map
\[ \{ P, Q \} \rightarrow \{u, v \}   \]
is not invertible. In other words, whereas $P$ and $Q$ generate the
full algebra $\cal B$ of operators in $\cal H$,
i.e. via Weyl formula
\[ A = \int f(s,t) \exp( i \frac{Qs + Pt}{\hbar}) \, \frac{ds dt}{2\pi \hbar}
\;\; , \]
it is not true for $u$ and $v$. In other words, algebra $\cal A$,
algebraically generated
by $u$ and $v$, is a proper subalgebra in $\cal B$.
Indeed, it is evident, that operators
\[ \hat{u} = e^{\frac{2\pi i Q}{q}}, \;\;\;\;
  \hat{v} = e^{\frac{2 \pi i P}{p}} \]
(and so all algebra, generated by them) commute with $u$ and $v$. The term
"algebraically generated" means that $\cal A$ is obtained by a suitable
closure of polinomials in $u$, $v$, $u^{-1}$ and $v^{-1}$. In this sense,
while $\hat{u}$ and $\hat{v}$ as elements of $\cal B$ are functions
of $u$ and $v$
\[ \hat{u} = v^{1/ \theta}, \;\;\;\;\; \hat{v} = u^{1/ \theta},  \]
they do not belong to $\cal A$ for generic $\theta$.

A natural question appears, what one is to add to $(u, v)$  to be able
to generate all $\cal B$. I want to argue, that in case when
$\theta$ is irrational, it is $(u, v, \hat{u}, \hat{v})$ which
generate all $\cal B$.

I came to the question of extention the algebra $\cal A$, generated by
$u$ and $v$, in course of work in collaboration with A.Volkov on
$U(1)$ lattice current algebra \cite{2}. This algebra is generated by
dynamical variables $w_n, n = 1, \ldots N$, with commutation relations
\[ w_n w_{n+1} = \omega w_{n+1} w_n   \]
\[  [ w_n, w_m ] =0 \;\;\;   |n-m| \geq 2  \]
and periodic boundary condition for finite chain. A natural
question about the finding of the operators $U$, realizing the shift
\[ w_n \rightarrow  w_{n+1},  \]
so that
\[ w_{n+1} = U w_n U^{-1}   \]
is easily reduced to the local problem in the Weyl algebra for a
pair of operators $u, v$ with relation
\[ uv = \omega vu \, ;   \]
one is to construct two operators $f$ and $g$ such that
\[   [f,u] = 0;   \;\;\;\; [g,v] = 0,  \]
\[ vfg = fgu.  \]
If we confine our search to algebra $\cal A$, generated by $u$ and $v$,
then we are to take
\[ g = r(v), \;\;\;  f = r(u),   \]
where $r(u)$ is an algebraic function of $u$
(we take into consideration the invariance with respect to the shift), and the
required property is achieved, if $r(u)$ is a solution of the
functional equation
\[ \frac{r(\omega u)}{r(u)} = \frac{1}{u}.           \]
This equation is easily solved by the series
\[ r(u) = \sum \omega^{\frac{n^2 - n}{2}} u^n     \]
which makes sense either if $|\omega| < 1$ or $\omega$ is a root of unity, when
we  can truncate the series. For arbitrary $\omega = e^{2 \pi i \theta}$ with
values on the circle, which is needed in applications, this answer
is unsatisfactory. The way out of this difficulty is proposed in my
Varenna lecture notes \cite{3} and corresponds exactly to the
extension of the algebra , where to look for the operators $f$ and $g$.
It was shown, that the operators $f$ and $g$
could be found in algebra generated by $u,v, \hat{u}, \hat{v}$. More on
this is below.

Now let us prove our main assertion.

{\bf Lemma} {\it Let $\theta$ be irrational. Then $u,v, \hat{u}, \hat{v}$
constitute an irreducible set of generators in $\cal B(H)$.}

It is enough to show that only unity commutes with our four generators.
We shall work in coordinate representation and realize the arbitrary
operator  $A$ as an integral one with kernel $A(x,y)$ (possibly a
generalized function).
\[ A \psi(x) = \int A(x,y) \psi (y) dy \; .  \]
Operators $u$ and $\hat{v}$ act as shifts
\[ u \psi (x) = \psi (x + q),    \;\;\;   \hat{v} \psi (x)  =
\psi ( x + \frac{2\pi \hbar}{p})  \; ,     \] and $v$ and $\hat{u}$
are multiplication operators
\[ v \psi (x) = e^{\frac{ipx}{\hbar}} \psi(x),     \;\;\;\;\;
 \hat{u} \psi (x) = e^{\frac{2 \pi i x}{q}} \psi (x) \; . \]
 Commutativity of $A$ with $\hat{u}$ shows, that
 \[ A(x,y) = \sum a_n (x) \delta (x-y -nq)    \]
 and then commutativity with $v$ leads to the condition
 \[ a_n (x) e^{\frac{i p x}{\hbar}} (e^{2 \pi i \theta n } -1) = 0.  \]
 If $\theta$ is irrational the last factor is not zero unless $n = 0$
 and we get
 \[ a_n (x) = 0, \;\;\;\; n \neq 0.   \]
 Thus $A$ is a multiplication operator
 \[ A(x,y) = a(x) \delta (x-y). \]
 Now commutativity with $u$ and $\hat{v}$ leads to condition of
 double periodicity of $a(x)$
\[ a(x) = a(x+q)  \, ,  \]
\[ a(x) = a(x + \frac{2\pi \hbar}{p}).     \]
The first condition shows that
\[ f(x) = \sum f_n e^{\frac{2 \pi i nx}{q}}   \]
and the second gives
\[ \sum  f_n e^{\frac{2 \pi i nx}{q}}  ( e^{\frac{2 \pi i n}{\theta}} -1) = 0
 \, , \]
from which due to irrationality of $1/\theta$ we see that
\[ f_n = 0, \;\;   n \neq 0.   \]
Finally we get
\[ A(x,y) = f_0 \delta(x-y),    \]
so that $A$ is a multiple of unity, which proves the Lemma.

For illustration I present a proper form for operators $f$ and $g$ introduced
above which was
obtained in \cite{3}. The answer looks as follows. Let us use $P$ and
$Q$ as generators of $\cal B$. Operators $f$ and $g$, defining the shift,
are to be looked for as functions of $P$ and $Q$ respectively
\[ f = r(P), \;\; g = r(Q).   \]
Function $r(P)$ satisfies  the functional equation
\[ \frac{r(P+p)}{r(P)} = e^{\frac{-i Pq}{\hbar}} e^{-i \pi \theta} \]
with a solution
\[ r(P) = \exp{ \frac{- \pi i \theta P^2}{p^2}}.      \]
It is not evident, that this $r(P)$ is generated by $u$ and $\hat{v}$
as it should. However the following is true.

Consider a more complicated functional equation
\[ \frac{s(P+p)}{s(P)} = \frac{1}{1+e^{\frac{ i Pq}{\hbar}}}.   \]
The solution is given by the integral
\[ s(P) = \exp{ \frac{1}{4} \int_{-\infty}^{\infty}
\frac{e^{\frac{Pq}{\hbar}\xi}}{\sinh{\pi \xi} \sinh{\pi \theta \xi}} \;
\frac{d \xi}{\xi}},            \]
where the singularity at $\xi = 0$ is circled from above. The calculation
of the integral by residues leads to the expression
\[ s(P) = \frac{s_{\theta} (u)}{s_{\hat{\theta}} (\hat{v})}     \]
where
\[  \hat{\theta} = - \frac{1}{\theta}      \]
and
\[ s_{\theta} (u) = \exp{ \frac{1}{2i} \sum \frac{(-1) u^n}{n \sinh{\pi n
\theta}} }. \]
For irrational real $\theta$ the series in the exponent suffers from the
small  denominators, however the ratio $s(P)$ is well defined. We see
that $s(P)$ is an element in algebra generated by $u$ and $\hat{v}$.

Now we have
\[ e^{\frac{- \pi i \theta P^2}{p^2}} = const \, s(P) s(-P)   \]
Thus the operators $f$ and $g$, generating shift $u \rightarrow v$ can be
explicitly constructed after our extension of algebra as function of
$u,v, \hat{u}, \hat{v}$. In other words, the shift is an outer
automorphism of the algebra $A_{\theta}$.

Now we turn to the first item of the title, namely the modular group. It is
clear that $\hat{u}$ and $\hat{v}$ also constitute a Weyl pair
\[ \hat{u} \hat{v}  = e^{\frac{-2 \pi i }{\theta}} \hat{v}  \hat{u}  \]
The only difference  from the original algebra of $u$ and $v$ is the
change of parameter
\[ \theta \rightarrow - \frac{1}{\theta}    \]
resembling one of the transformations of the modular group, usually
called $S$. Another one, namely $T$
\[ \theta \rightarrow \theta +1   \]
does not change the algebra ${\cal A}_{\theta}$. Thus we have natural
action of $S$ and $T$ on generators $u$, $v$
\[ S(u) = \hat{u}, \;\;\; S(v) = \hat{v}, \]
\[  T(u) = u ,     \;\;\;  T(v) = v.     \]
It continues further as follows
\[ S(\hat{u}) = u, \;\;\; S(\hat{v}) = v, \]
\[  T(\hat{u}) = \hat{u}^{-1}u,  \;\;\;  T(\hat{v}) = \hat{v}^{-1} v. \]
It is easy to check, that the relations of modular group are satisfied
in the following form
\[ S^2 = {\rm id}; \;\; (TS)^3 = \sigma     \]
where $\sigma$ is an automorphism of $A_{\theta}$
\[ \sigma \left( \begin{array}{c}  u  \\
                                   v
                \end{array}  \right)    =
          \left( \begin{array}{c}  u^{-1} \\
                                   v^{-1}
                \end{array}  \right).               \]
and the same for $\hat{u}$ and $\hat{v}$. Thus we see that with any
irrational $\theta$ we can associate the action of the modular
group on the algebra $\cal B$.

Pair of algebras $A_{\theta}$ and $A_{-1/\theta}$ was considered by
A.Connes in a different context \cite{4}. He commented
that $A_{\theta}$ (and $A_{-1/ \theta}$)  is a factor ${\rm II}_1$.
Indeed, one can introduce a trace over $A_{\theta}$ in the following manner.
For general element
\[ a = \sum a_{mn} u^m v^n     \]
we have
\[ tr a = a_{00}     \]
It is easy to check that
\[ tr (ab) = tr (ba) \]
so that the main property of trace is satisfied. For irrational $\theta$
algebra ${\cal A}_\theta$ is infinite dimensional factor in $\cal B$.
However, $tr(\rm{I})=1$. That is why it is factor ${\rm II}_1$.

Our assertion shows that
for irrational $\theta$ factors $A_{\theta}$ and $A_{-1/\theta}$ are commutants
of
each other in $\cal B$ and together they generate $\cal B$.
In other words we see that one degree of freedom is virtually
divided into two! Whereas this comment is irrelevant for one
particle quantum mechanics, it could be important in applicaions to quantum
field theory, where the Weyl-type operators appear in lattice
regularizations.

Indeed, in such a regularization one uses the exponents of canonical
fields and usually speaks of the "compactified" fields with $\theta$
being called a "radius of compactification" (see, e.g., \cite{5}).
 Operators a-la $u$ and
$v$ ( or $w_n, \; n= 1, \ldots$ ) serve as generators of these
"compactified" degrees of freedom. The dual operators a-la $\hat{u}$
and $\hat{v}$ are to be interpreted as generators of additional
degrees of freedom, called "winding" modes or solitonic modes etc.

Radii of compactification play the role of the coupling constants with
the usual solitonic feature \cite{6}: weak interaction of main modes
corresponds to strong interaction of solitonic modes.

I believe that naive mathematics presented above bears the essential
explanation of the necessity of taking into account the solitonic modes
and appearence of duality and/or modular covariance in dynamical problems.

This work was partly supported by ISF grant N R2H000 and grant of
Finnish Academy.

It is a pleasure to acknowledge the constructive discussions with
A.Connes, M.Flato, A.Morozov, A.Niemi, D.Sternheimer and fruitful
collaboration with  A.Volkov.

\end{document}